\documentclass{sig-alternate}
\usepackage{array}
\usepackage{graphicx}

\begin{document}
\title{An experimental evaluation of de-identification tools for electronic health records}

\vspace{5mm}
\numberofauthors{2}

\author{
\alignauthor Jie Qian\\
  \affaddr{International Institute for Software Technology}\\
  \affaddr{United Nations University}\\
  \affaddr{Macao SAR, China}\\
  \email{qj@iist.unu.edu}
\alignauthor Nafees Qamar\\
  \affaddr{International Institute for Software Technology}\\
  \affaddr{United Nations University}\\
  \affaddr{Macao SAR, China}\\
  \email{nqamar@iist.unu.edu}
  }
  
\maketitle

\begin{abstract}
The robust development of Electronic Health Records (EHRs) causes a significant growth in sharing EHRs for clinical research. However, such a sharing makes it difficult to protect patients' privacy. A number of automated de-identification tools have been developed to reduce the re-identification risk of published data, while preserving its statistical meanings. In this paper, we focus on the experimental evaluation of existing automated de-identification tools, as applied to our EHR database, to assess which tool performs better with each quasi-identifiers defined in our paper. Performance of each tool is analyzed \textit{wrt.} two aspects: individual disclosure risk and information loss. Through this experiment, the generalization method has better performance on reducing risk and lower degree of information loss than suppression, which validates it as more appropriate de-identification technique for EHR databases.
\end{abstract}


\section{Introduction}

For the purpose of effective health care, sharing patient's EHR is one of the trends empowering modern health information systems (HIS); and with standardized EHR specifications like HL7 and openEHR\cite{openEHR}, sharable EHR system is at the edge of practice. Huge amounts of patients' EHRs would then be used by numerous clinical researchers and online diagnosis services. However, if these EHRs are directly published to the public, it would inevitably lead to serious confidentiality problems. In reality, hospitals have confidentiality agreements with patients, which strictly forbid HIS discloses any identifiable information on individuals. In addition to that, laws such as HIPAA\cite{HIPAA} explicitly state the confidentiality protection on health information, where any sharable EHR system must legally comply with.

One approach on confidentiality protection is to remove any identifiable information (i.e., patient's name, SSN, etc) of an EHR. However, adversary can still re-identify a patient by inferring from external information. A research \cite{simp} indicates that 87 percents of the population of U.S. can be distinguished by sex, date of birth and zip code. Such a combination of attributes, which can uniquely identify individual, is defined as quasi-identifiers. If adversary has acknowledged of these quasi-identifiers, an attacker might recognize an individual and take advantage of these clinical data. On the other hand, we can find out most of these quasi-identifiers have statistical meanings in clinical researches. Thus, there exists a paradox between reducing the likelihood of disclosure risk and retaining the data quality. For instance, if any information of patient's residence is excluded from the EHR, it would disable related clinical partners to catch the spread of a disease. Conversely, releasing data including total information of patient's residence, sex and date of birth would bring a higher disclosure risk.

De-identification is defined in \cite{HIPAA} as a technology to remove the identifying information such as name, SSN from the published dataset. Specifically, it deals with the challenge mentioned above by protecting the data under a maximum tolerable disclosure risk while preserving the data of an acceptable quality. In recent years, several typical privacy criteria (i.e., \textit{k}-anonymity \cite{kanony}, \textit{l}-diversity \cite{ldiver}, and \textit{t}-closeness \cite{tclose}) and anonymization methods (i.e., generalization, suppression and etc) have been proposed. Based on these works, several de-identification tools (i.e., CAT, $\mu$-Argus and sdcMicro) are developed to manage disclosure risk. Each tool has its sample demonstration and some of them have been applied on real datasets \cite{cvt}. So far, some researchers \cite{pre} \cite{engine} have evaluated some of these tools. In \cite{pre}, it focus on the technical details of the anonymization process and methods. However, it does not present any practice on these tools. In \cite{engine}, it evaluates a comprehensive de-identification engine using a dataset of surgical pathology reports. It analyzes the anonymization steps optimized by the engine, but not in a systemic way. Overall, none of these work provides an insight about the best tool and method for de-identifying pubished dataset. Our study focuses on a systemic way to compare the existing tools based on experimental evaluation.

We propose an experiment on our EHR database to evaluate the performance of each de-identification tool. Then we find the most suitable tool for releasing EHRs by judging the capability of minimizing data disclosure risk and the distortion of the results.

The paper is organized as follows. In Section~\ref{thetools} briefly introduces the experimented tools; Section~\ref{ehrdb} analyses the EHR database and then lists the potential quasi-identifiers; Section~\ref{methods} introduces the design of experiment; Section~\ref{results} presents the results of experiment; Section~\ref{conclusion} presents the conclusion of the experiment and discusses the limitation.

\section{The experimented tools}\label{thetools}

\begin{table*}[t]
\begin{center}
\caption{Featuring the three de-identification tools}
\begin{tabular}{|l|l|l|m{3cm}|m{3cm}|}
\hline\textbf{}
\textbf{Tools}&\textbf{Input Data}&\textbf{Privacy Criterion}&\textbf{Anonymization Approach}&\textbf{Data Evaluation}\\
\hline
CAT&Meta and microdata&\textit{l}-diversity, \textit{t}-closeness
&Generalization, Suppression&Comparison, Risk analysis\\
\hline
$\mu$-Argus&Meta and microdata&\textit{k}-anonymity&Global recoding, Local Suppression, etc&Risk analysis\\
\hline
sdcMicro&Database&\textit{k}-anonymity&Global recoding, Local Suppression, etc&Comparison, Risk analysis\\
\hline
\end{tabular}
\end{center}
\end{table*}

A number of groups \cite{uarg}\cite{sdc}\cite{cat} are actively developing their de-identification tools, aiming to enable users publish safer data. They have adopted different approaches which reflect their particular interests and expertise. However, all these tools include the same anonymization process in which iteratively approximates a privacy criterion.

\subsubsection*{CAT}\cite{cat}
is developed by a database group at Cornell University. It anonymizes data using generalization technique \cite{gen}, specifically, replacing the values of quasi-identifiers into value ranges. It provides graphical interface, which eases the operation like adjusting the settings of privacy criterion or checking the current disclosure risk. In terms of usability it presents the contingency tables and density graphs between the original and anonymous data, which gives users an intuitive way to learn the information loss during the de-identification process.

\subsubsection*{$\mu$-Argus}\cite{uarg}
is part of the CASC project, which is partly sponsored by the European Union. The main part of this software is developed and fully tested at Statistics Netherlands. In particular, it supports almost all the typical de-identification approaches (i.e., global recoding, local suppression, PARM, etc) which enable a variety of selections to enhance security.

\subsubsection*{sdcMicro}\cite{sdc}
is developed by Statistics Austria based on R, a highly extensive system for statistical computing. It contains almost the same classic anonymization methods as $\mu$-Argus. Since R can be seen as a function and class-oriented programming language, it offers a facility for designing and writing functions for particular research purposes.

Table 1 illustrates a preliminary summary of the similarities and differences of these tools, allowing an security specialist to have a better intuition of the techniques behind.

\section{EHR Database}\label{ehrdb}

In cooperation with the dialysis center of Kiang Whu Hospital, Macau, we have implemented a system for utilizing its EHR management. The test database consists of 1000 EHR samples in which a total of 183 variables have been recorded.

De-identifying such a database is a challenge since a large amount of sensitive clinical data can be involved in any common request. Supposing an organization request for a published dataset on patients' infectious disease histories, the corresponding quasi-identifiers undergo the potential risk of leaking patient's privacy. An adversary could determine one of the quasi-identifiers referenced to a female born on 12/04/64, sent to Kiang Whu Hospital last Friday, and living in Taipa is exactly his neighbor. Then he could find his neighbor has an infectious disease history of HCV.

Here, we consider a subset of the combination of the following variables in the database: gender, date of birth, place of birth, province of residence, and zip code as a quasi-identifier. Throughout this paper, we use the following notations: QID = quasi-identifier, ZC = zip code, DoB = date of birth, YoB = year of birth, DoR = district of residence, PoB = place of birth.

For each quasi-identifier, we counted the number of distinct values in the database, which indicates the number of anonymity sets; the number of patients sharing a specific value that represents the anonymity set size \textit{k}. We chose quartiles as a means to indicate the value distribution of the anonymity sets.

\begin{table*}[t]
\begin{center}
\caption{Anonymity set size \textit{k} for various quasi-identifiers}
\begin{tabular}{|l|l|l|l|l|l|l|}
\hline
\textbf{QID:}& \textbf{numbers of sets}&\textbf{Min.}&\textbf{1st Qu.}&\textbf{Median}&\textbf{3rd Qu.}&\textbf{Max}\\
\hline
ZC&38&9&20&25&31&51\\
\hline
ZC+gender&76&2&10&13&16&30\\
\hline
ZC+DoB&997&1&1&1&1&2\\
\hline
ZC+YoB&659&1&1&1&2&5\\
\hline
ZC+PoB&280&1&1&1&2&37\\
\hline
ZC+gender+YoB&804&1&1&1&1&4\\
\hline
ZC+gender+PoB&341&1&1&1&2&22\\
\hline
gender+DoB&998&1&1&1&1&2\\
\hline
gender+YoB&70&5&10&13&19&38\\
\hline
gender+DoR&14&55&62&72&77&91\\
\hline
gender+PoB&44&2&5&7&9&369\\
\hline
gender+DoR+PoB&191&1&1&2&2&67\\
\hline
gender+PoB+YoB&336&1&1&1&2&31\\
\hline
gender+DoR+YoB&398&1&1&2&3&11\\
\hline
gender+DoR+PoB+YoB&638&1&1&1&2&9\\
\hline
\end{tabular}
\vspace{-5mm}
\end{center}
\end{table*}

Table 2 shows the statistical characteristics of anonymity set size \textit{k} for various quasi-identifiers. The second column indicates the number of anonymity sets in our database for a given quasi-identifier. Generally, during the de-identification process, the larger the number of distinct anonymity sets, the less information distortion on the published dataset, because the anonymity set tends to be smaller in that case and removing one affects little on the overall dataset. The min and max values denote the smallest and largest anonymity set.

According to Table 2, it is clear that some quasi-identifiers lead to particularly high disclosure risks, because more than half of their anonymity sets are smaller than 2, which means a large portion of patients can be unambiguously identifiable by that quasi-identifier. For instance, for \{ZC+DoB\}, we can find that '\textit{k}=1' is up to the 3rd quartile, which means at least 75 percents of the patients are unambiguously identifiable by zip code and date of birth. Also, some quasi-identifiers are weaker because their smallest anonymity set is more than 5, such as \{ZC\}, \{gender+DoR\} and \{gender+YoB\}. Overall, it turns out quasi-identifier that contains date of birth, place of birth and year of birth are most identifiable.

We also found that the size of anonymity sets for which quasi-identifiers containing place of birth has a significant increase between the third quartiles and max value. It means a relatively large group of patients converge to one characteristic. This is because most of the patients of Kiang Whu hospital are Macau citizens. Consequently, patients who were born elsewhere are of sparse distribution and more likely to be unambiguously identifiable by their \{gender+PoB\} or \{ZC+PoB\}. Table 2 also clearly shows that year of birth, a reduction of date of birth, increases the de-identifiability: the median anonymity set size for \{gender+YoB\} is 13, whereas for \{gender+DoB\} is only 1.

Table 3 shows the actual number of patients that belongs to those anonymity sets, for example, for \{ZC+DoB\}, only two patients can be found in anonymity sets that have \textit{k}$\leq$5. The larger the value in the columns '\textit{k}=1' and '\textit{k}$\leq$5', the larger the portion of the patients that is covered by anonymity sets of small sizes, and the stronger the quasi-identifier identify patients. The number indicates that \{ZC+DoB\} is the strongest quasi-identifier, because almost all patients have \textit{k}=1. However, zip code alone is a weaker quasi-identifier, because none of patients is in the first two columns.

Similarly, \{gender+DoB\} is a very strong quasi-identifier mainly because date of birth poses a significant privacy risk for nearly all the patients in our database. In this experiment, we replaced date of birth to year of birth before the experiment.

The numbers for \{ZC+gender+YoB\} indicates that 63.7 percents of the patients can be unambiguously identified by this quasi-identifier. For \{gender+DoR+PoB+YoB\}, it shows that nearly half of the patients can be unambiguously identified.

\begin{table}
\begin{center}
\caption{Number of EHR data per anonymity set size, for various quasi-identifiers}
\begin{tabular}{|l|l|l|l|l|}
\hline
\textbf{QID:}&\textbf{\textit{k}=1}&\textbf{\textit{k}$\leq$5}&\textbf{\textit{k}$\leq$10}&\textbf{\textit{k}$\leq$50}\\
\hline
ZC&0&0&9&949\\
\hline
ZC+gender&0&2&179&1000\\
\hline
ZC+DoB&994&1000&1000&1000\\
\hline
ZC+YoB&418&1000&1000&1000\\
\hline
ZC+PoB&199&294&309&1000\\
\hline
ZC+gender+YoB&637&1000&1000&1000\\
\hline
ZC+gender+PoB&237&333&664&1000\\
\hline
gender+DoB&994&1000&1000&1000\\
\hline
gender+YoB&0&10&188&1000\\
\hline
gender+DoR&0&0&0&0\\
\hline
gender+PoB&0&57&242&294\\
\hline
gender+DoR+PoB&90&294&304&542\\
\hline
gender+PoB+YoB&240&354&575&1000\\
\hline
gender+DoR+YoB&134&864&989&1000\\
\hline
gender+DoR+PoB+YoB&435&958&1000&1000\\
\hline
\end{tabular}
\vspace{-5mm}
\end{center}
\end{table}

\section{Methods}\label{methods}

In order to assess the performance of the selected de-identification tools with our EHR database, we designed our experiment of the following four aspects.

\subsubsection*{Selection of quasi-identifiers}
Judging from Table 3, we found \{ZC+gender
+YoB\} (denoted as $QID^{1}$) and \{gender+DoR+PoB+YoB\} (denoted as $QID^{2}$) are the most representative quasi-identifiers for this database (note that we excluded the quasi-identifiers that contain date of birth).

\subsubsection*{Selection of privacy criteria}
To ease the comparison of the tools, we provided \textit{k}-anonymity for this dataset. In this experiment, we set \textit{k} to 2, which means the minimum value of anonymity set size that is safe for $QID^{1}$ and $QID^{2}$.

\subsubsection*{Dimensions of comparison}
Two dimensions of comparison were identified. The first dimension is the individual disclosure risk of the published datasets regarding the above quasi-identifiers. An accurate measure in terms of the individual risk on a quasi-identifier was defined as the following formula in \cite{book}.

\begin{equation}\label{formula 1}
\xi = \frac{1}{n}\sum_{k=1}^{K}f_{k}r_{k}
\end{equation}

For a quasi-identifier, $f_{k}$ denotes the size of k-th anonymity set of the dataset;  $r_{k}$ denotes the probability of re-identifi\-cation of a k-th anonymity set; \textit{n} denotes the total number of the records. A higher number indicates that the published dataset undergo a higher probability of disclosing patient's privacy. Generally, individual disclosure risk is related to the threshold value. Suppose that a threshold $r^{*}$ has been set on the individual risk (see formula (1)), unsafe records are those for which $r_{k}$$\leq$$r^{*}$. When threshold value set to 0.5, it ensures the dataset to achieve 2-anonymity. Similarly, when set to 0.2, it requires the dataset to achieve 5-anonymity.

The second dimension is the information loss for the published datasets. A strict evaluation of information loss must be based on a comparison between original dataset and published dataset. A metric called \textit{Prec} was proposed by Sweeny in \cite{inmetric}. For each quasi-identifier, \textit{Prec} counts the ratio of the practical height applied to the total height of the generalization hierarchy. Consequently, the more the variables are generalized, the higher the information loss. However, \textit{Prec} has been criticized not considering the size of the generalized cells. Also, it doesn't account for the information loss caused by suppression method. Another commonly used metric is DM* \cite{optik}, which addresses on the weakness of \textit{Prec}. But it has also been criticized by \cite{dm} because it does not give intuitive results when the distributions of the variables are non-uniform. Therefore, these two metrics are not suitable for this experiment.

The distribution of anonymity set size for a quasi-identifier indeed represents the risk model of a dataset, where is feasible to compare the original dataset and published dataset. Looking into Table 2, it is clear that the anonymity set size of \{gender+PoB\} (denoted as $QID^{3}$) and \{gender+DoR+PoB\} (denoted as $QID^{4}$) has significant increase between the third and forth quartile than other quasi-identifiers. In other words, the individual disclosure risk has a significant decrease in the third and fourth quartile, which means that quasi-identifier is more sensitive to a change in its anonymity sets. To simplify the results, we measure the information loss in terms of the slope of anonymity set size for each $QID^{3}$ and $QID^{4}$ in the third and fourth quartile.
The information loss for a quasi-identifier is:

\begin{equation}\label{formula 2}
\lambda=\frac{\frac{\partial{R^{'}}}{\partial{Num^{'}}}}{\frac{\partial{R}}{\partial{Num}}}= \frac{\frac{k(n+1)^{'}-k(n)^{'}}{Sum^{'}}}{\frac{k(n+1)-k(n)}{Sum}}
\end{equation}

Where k(n) represents the anonymity set size at the n-th quartile, and \textit{Sum} represents the number of distinct anonymity sets in the dataset. The above formula usually yields a positive value.  A higher number suggests a higher information loss of the original dataset.

\subsubsection*{Principal methods used for de-identification}
Although different methods for acquiring \textit{k}-anonymity criterion have been implemented in these tools, we present here a broad classification depending on the main techniques used to de-identify quasi-identifers. Specifically, we classify anonymization methods in two categories as follows: generialization and suppression, as proposed in \cite{met}. Other methods, which randomly replaces the values of quasi-identifiers (e.g, adding noise), distort the individual data in ways that sometimes results in incorrect clinical inferences. As these methods tend to have a low acceptance among clinical researchers, we decided not to apply them to the EHR database.

Generalization provides a feasible solution to achieving \textit{k}-anonimity by transforming the values in a variable to the optimized value ranges reference to the user-defined hierarchies. Particularly, global recording means that generalization performed on the quasi-identifiers across all of the records, which ensures all the records have the same recoding for each variable.

Suppression means the removal of values from data. There are three general approaches to suppression: casewise deletion, quasi-identifier removal, and local cell suppression, where CAT applied the first approach; $\mu$-argus and the sdcMicro applied the third approach. For the same affected number of records, casewise deletion always has a higher degree of distortion on the dataset than local cell suppression. In most case, suppression leads to less information loss than generalization because the former affects single records whereas the latter affects all the records in the dataset. However, the negative effect of missing values should be considered.

\section{Results}\label{results}
Before starting our experiment, we indexed our EHR data\-base into microdata and metadata. For instance, we mapped 7 identifiable variables in \textbf{PatientRecord} table to categorical variables, 1 to numerical variables, 9 to string variables and removed 18 variables that were either illegal to release(i.e., patient's name, SSN) or irrelevant to research purpose (i.e., time stamp, barcode).We also truncated the value of date of birth variable into year of birth.

\begin{figure}[!t]
\begin{center}
\includegraphics[width=2.7in]{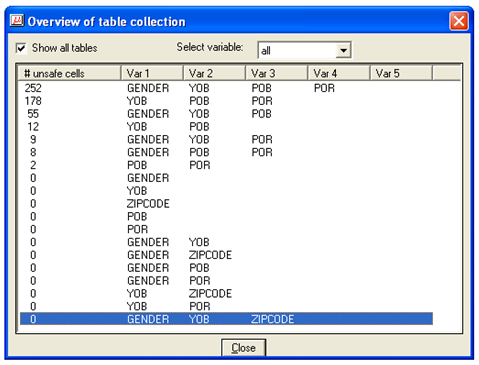}
\caption{An overview of unsafe records for various quasi-identifiers}
\label{fig:1}
\end{center}
\end{figure}

We started by anonymizing our dataset using $\mu$-Argus. First, we specified the combination of variables to be inspected as $QID^{1}$ and $QID^{2}$ with the threshold set to 1(maximum value of anonymity set size \textit{k}, which is considered unsafe). It should be noted that the individual risk model was restricted in $\mu$-Argus because there was an overlap between the quasi-identifiers. Then the tool counted the number of the unsafe records that are unambiguously identifiable for each combination of variables. As suggested in the user's manual, the first anonymization method we applied was global recoding. Specifically, 22 different values in place of birth variable were equivalently generalized to 8 categories; 35 different values in year of birth variable were generalized to 12 categories; the last digit of zip code was removed. As shown in Figure~\ref{fig:1}, the number of unsafe records decreased from 637 to 0 and 435 to 252, respectively for $QID^{1}$ and  $QID^{2}$. It is clear that global recoding significantly decreases the risk of re-identification on  $QID^{1}$. However, for  $QID^{2}$, 252 out of 1000 patients remain to be unambiguously identifiable.

After dealing with categorical variables, we found that micro aggregation method was not practical, because the minimum frequency of the numeric variable is far above the minimum requirement for safe anonymity set size. Then we applied local suppression method to protect the remaining unsafe records. This led to 75 values in gender variables and 121 values in place of birth variables suppressed from the dataset.

Following we started sdcMicro. Due to the character encoding issue on ODBC, we collated our dataset from Traditional Chinese to UTF-8, which resulted in character loss on some of the values in place of birth and district of residence variables. Then we used freqCalc function in sdcMicro to calculate the number of unsafe records for $QID^{2}$. The result shows that 411 records could be unambiguously identified by  $QID^{2}$, contrast to 435 in Table 3, which indicates an inaccuracy deviation of 5.5\% on  $QID^{2}$.

Similarly, we first applied the sdcMicro function globalRecode to the dataset. It turns out year of birth variable generalized to the same 12 categories, which reduced the number of unsafe records to 244 and 254, respectively for $QID^{1}$ and $QID^{2}$.

Then the function localSupp could be used to apply local suppression method. Using the threshold value of 0.5 (to achieve 2-anonymity as mentioned in Section IV), localSupp was first applied to $QID^{1}$. This led to a suppression of 244 values in zip code variable and 20 values in year of birth variable. Again, calculating the number of unsafe records for this quasi-identifier, we found that the published dataset reached 4-anonymity and the maximum value of individual risk decreased to 0.143. For $QID^{2}$, we notice that most of the unsafe records have re-identification risk over 0.89. With the threshold value to 0.89, suppression of 254 values in place of birth variable were done. We observed only 3 records with anonymity set size \textit{k}=1. Then suppression (threshold value = 0.5) was applied, 3 values in district of residence variable were suppressed.

\begin{figure}
\begin{center}
\includegraphics[width=3in]{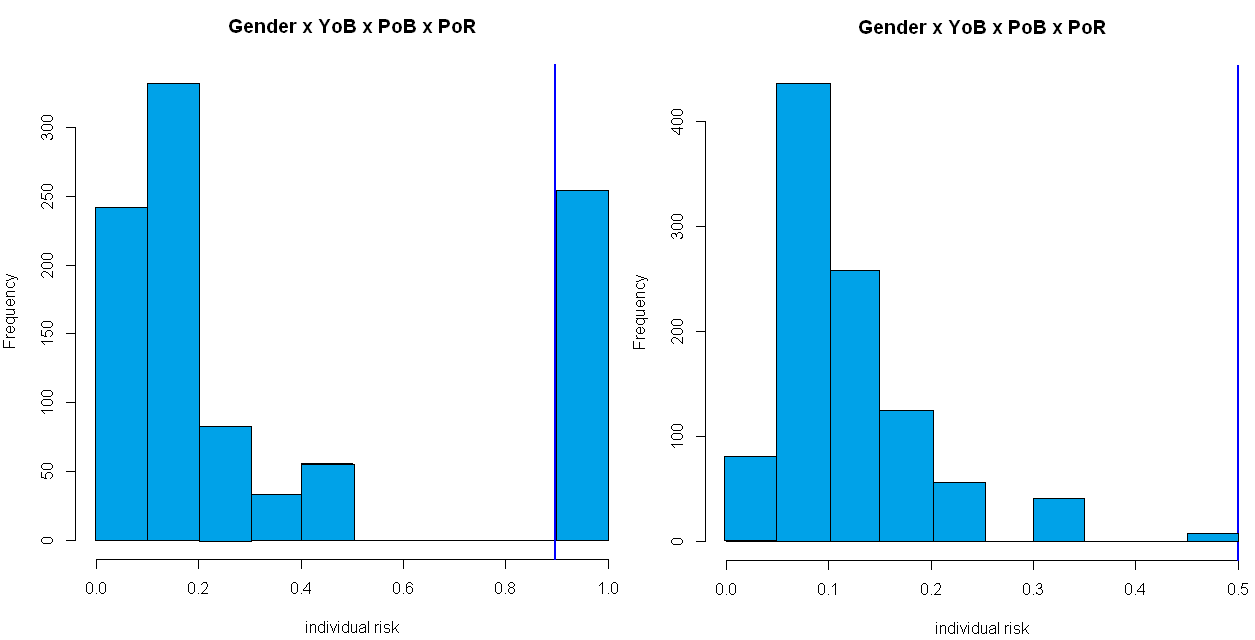}
\caption{The individual risk of the dataset for $QID^{2}$}
\label{fig:2}
\end{center}
\end{figure}

The left side of Figure~\ref{fig:2} shows the distribution of individual risk of the original dataset for $QID^{2}$, while the right side shows the result of the published dataset. It is clear that the maximum value of individual risk decreased from 1.0 to 0.5. After three suppressions were done, for each quasi-identifier, the dataset satisfied 2-anonymity.

The third tool was CAT. As the tool restricts one quasi-identifier per anonymization process, we specified two quasi-identifiers respectively. Since CAT doesn't provide k-anony\-mity directly, we choose \textit{t}-closeness criteria instead. We first provided \textit{t}-closeness criteria on the $QID^{1}$ with a threshold value \textit{t} to 0.5, which means the maximum value of individual disclosure risk is 0.5. This led to generalization method applied to year of birth and zip code variables. Specifically, every ten values in zip code variable were generalized into one category, which addressed the same effect on the published dataset as a truncation of the last digit of this variable; every two values in year of birth variable were generalized into one category.

\begin{figure}
\begin{center}
\includegraphics[width=3.3in]{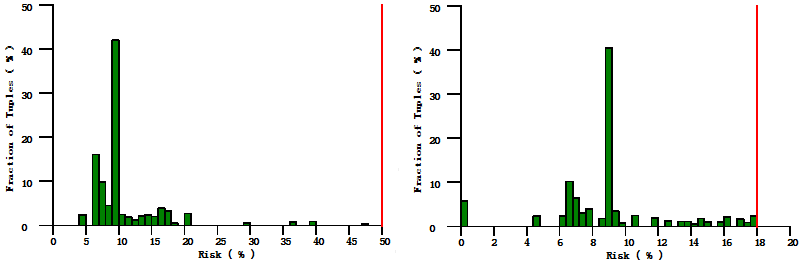}
\caption{The individual risk of the dataset for $QID^{1}$}
\label{fig:3}
\end{center}
\end{figure}

As the left side of Figure~\ref{fig:3} shows, the current maximum value of individual disclosure risk is 0.5. After deleting 58 records, the maximum value decreased to 0.18. Looking into the right side of Figure~\ref{fig:3}, which presents the distribution of individual disclosure risk of $QID^{1}$ on the published dataset, we found that less than 20 percents of the records have the risk above 0.1 and 2-anonymity was reached.

We then provided the \textit{t}-closeness criteria with a threshold value \textit{t} set to 0.978 on $QID^{2}$. This led to generalization method applied on year of birth variable, place of residence variable and place of birth variable. In particular, the values in place of residence variable were mapped into one category. The values in year of birth variable were equivalently mapped to 12 categories. The values in place of birth variable were equivalently mapped to 5 categories. After removing 57 records, the maximum value of individual risk decreased to 0.15.

For each quasi-identifier, these de-identification tools were able to publish the EHR dataset that satisfy 2-anonymity. We then analyzed the published datasets in terms of two aspects: individual disclosure risk and information loss.

\begin{table}
\begin{center}
\caption{Two dimensions of comparison for various quasi-identifiers}
\begin{tabular}{|l|l|l|l|}
\hline
\textbf{Maximum level}&\textbf{CAT}&\textbf{sdcMicro}&\textbf{$\mu$-Argus}\\
\hline
 $\xi$ for $QID^{1}$ &0.149&0.143&0\\
\hline
 $\xi$ for $QID^{2}$ &0.402&0.500&0.384\\
\hline
 $\lambda$ for $QID^{3}$ &3.600&2.028&1.682\\
\hline
 $\lambda$ for $QID^{4}$&86.631&3.261&1.783\\
\hline
\end{tabular}
\vspace{-5mm}
\end{center}
\end{table}

Here we calculate the individual disclosure risk $\xi$ of all published dataset using formula (1). Table 4 indicates that $\mu$-Argus has produced safer dataset than the others because it could protect patient's privacy under the lowest maximum individual risk for both quasi-identifiers. In particular, all records in the dataset produced by $\mu$-Argus satisfy 2-anonymity for $QID^{1}$. Following, we evaluated each published dataset in terms of their information loss $\lambda$ (see formula (2)). Since CAT generalized all the values in the place of residence variable into one category, it led to a significant information distortion on $QID^{4}$. In contrast to CAT and sdcMicro, $\mu$-Argus takes the lowest information loss for both quasi-identifiers to reach 2-anonymity.

\section{Conclusion and Discussion}\label{conclusion}

Our study compared three de-identification tools that are available for researchers to publish dataset using anonymization techniques. We first introduced the features of each tools, the anonymization methods behind, and the privacy criteria adopted.  Following, we analyzed the EHR database in terms of two categories: anonymity set size \textit{k} and number of EHR data per anonymity set size for 15 quasi-identifiers. We found quasi-identifiers that contain place of birth variable and year of birth variable are most identifiable. Then we selected two quasi-identifiers to be observed and anonym\-ized. We also include two formulas, based on which the published dataset of each tool can be examined in two dimensions: individual disclosure risk and information loss. For each tool, we outlined the anonymization process and provided 2-anonymity. Finally, we calculated the information loss and individual risk of each published dataset. As $\mu$-Argus produced the safest records and caused the lowest information loss among these tools, it is the most suitable de-identification tool for anonymizing our EHR database.

Numerical methods are proposed to anonymizing quasi-identifiers from disclosing individual's sensitive information. However, some methods such as masking, sampling were not implemented in these de-identification tools. Therefore, we are not able to evaluate the effectiveness of these methods on our EHR database.

Besides the results show the performance, it indicates the difference of each tool on the algorithm of optimizing generalization steps. For instance, 254 values in place of birth variable were suppressed in sdcMicro, while all the values were generalized to 8 categories in $\mu$-Argus. As $\mu$-Argus generalized more variables than sdcMicro, it benefits less records being suppressed, and the statistical meanings of these variables can be preserved. This also shows a specialty on our experiment that generalization causes a lower information loss than suppression when the latter takes certain percents of the total records. Consequently, as applied to our EHR database, generalization method are more suitable than suppression method.

For the purpose of comparison, we consider \textit{k}-anonymity as the only privacy criteria, which might lead to attribute disclosure problem on patient's clinical data. Since there is no de-identification approach applied to clinical variables (i.e. infectious disease, blood type in \textbf{PatientRecord} table), an attacker can discover a patient's clinical information when there is a little diversity in those clinical variables. Such problems are planned to be solved in the future development of de-identification component of our project.

\section{Acknowledgments}

This work has been supported by the project SAFEHR funded by Macao Science and Technology Development Fund.

\bibliographystyle{abbrv}
\bibliography{reference}

\end{document}